\documentclass[sigconf,author version,nonacm]{acmart}
\AtBeginDocument{%
  \providecommand\BibTeX{{%
    \normalfont B\kern-0.5em{\scshape i\kern-0.25em b}\kern-0.8em\TeX}}}

\copyrightyear{2024} 
\acmYear{2024} 
\setcopyright{acmlicensed}\acmConference[GLSVLSI '24]{Great Lakes Symposium on VLSI 2024}{June 12--14, 2024}{Clearwater, FL, USA}
\acmBooktitle{Great Lakes Symposium on VLSI 2024 (GLSVLSI '24), June 12--14, 2024, Clearwater, FL, USA}
\acmDOI{10.1145/3649476.3660389}
\acmISBN{979-8-4007-0605-9/24/06}

\begin{document}

\title{
Designing Reconfigurable Interconnection Network of Heterogeneous Chiplets Using Kalman Filter}

\author{Siamak Biglari}
\authornote{Both authors contributed equally to this research.}
\email{siamakbiglariardebili@my.unt.edu}
\affiliation{%
  \institution{University of North Texas, USA}
  \country{}
}
\author{Ruixiao Huang}
\authornotemark[1]
\email{ruixiaohuang@my.unt.edu}
\affiliation{%
  \institution{University of North Texas, USA}
  \country{}
}
\author{Hui Zhao}
\email{hui.zhao@unt.edu}
\affiliation{%
  \institution{University of North Texas, USA}
  \country{}
}
\author{Saraju Mohanty}
\email{saraju.mohanty@unt.edu}
\affiliation{%
  \institution{University of North Texas, USA}
  \country{}
}

\begin{abstract}
  Heterogeneous chiplets have been proposed for accelerating high-performance computing tasks. Integrated inside one package, CPU and GPU chiplets can share a common interconnection network that can be implemented through the interposer. However, CPU and GPU applications have very different traffic patterns in general. Without effective management of the network resource, some chiplets can suffer significant performance degradation because the network bandwidth is taken away by communication-intensive applications. Therefore, techniques need to be developed to effectively manage the shared network resources. In a chiplet-based system, resource management needs to not only react in real-time but also be cost-efficient. In this work, we propose a reconfigurable network architecture, leveraging Kalman Filter to make accurate predictions on network resources needed by the applications and then adaptively change the resource allocation. Using our design, the network bandwidth can be fairly allocated to avoid starvation or performance degradation. Our evaluation results show that the proposed reconfigurable interconnection network can dynamically react to the changes in traffic demand of the chiplets and improve the system performance with low cost and design complexity. 
\end{abstract}




\maketitle

\section{Introduction}

\begin{figure}
    \centering
    \scalebox{0.7}{\includegraphics[width=0.9\linewidth]{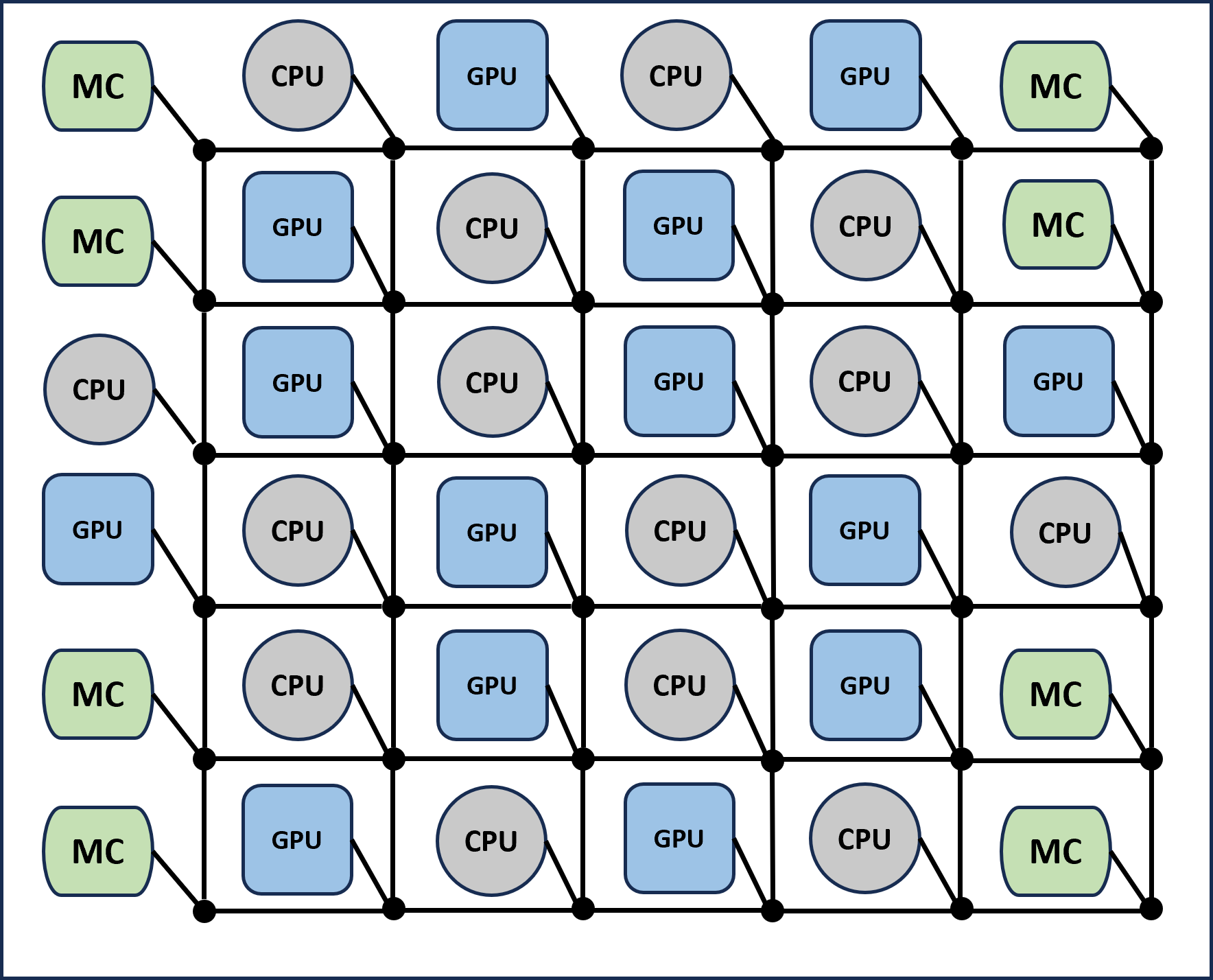}}
    \vspace{-0.2 cm}
 \caption{An example of a heterogeneous chiplet.}
    \label{fig:cpu-gpu}
    \vspace{-0.5 cm}
\end{figure}

Heterogeneous multi-core architectures, incorporating both CPUs and GPUs, have found many applications across a spectrum of computing platforms, including high-performance servers, mobile devices, electronic gadgets, desktop computers, and gaming consoles. In such multi-core systems, chips containing CPU or GPU cores are integrated into a single die such as Intel's Sandy Bridge \cite{Gwennap2010}, ARM's MALI \cite{Bratt2015}, and AMD's Fusion APUs\cite{Lee2013}\cite{lc-conflict-free}.
 Both types of cores can collaborate within a unified virtual address space and programming framework like CUDA and OpenCL. Recent research also proposed to use different virtual address spaces between CPUs and GPUs. However, merging distinct CPU and GPU cores into a same die introduced new challenges in the system design\cite{onur}.

Chips are usually designed for specific targets, making them less scalable in addressing diverse resource requirements and thus becoming less cost-efficient when integrating them. In fact, certain types of cores can be separately manufactured to meet their specific requirements, potentially reducing the cost and improving the yields. As a solution to heterogeneous system integration, chiplets have been proposed to tackle the rising cost and scalability issue. In a chiplet system, the single die is broken into multiple smaller chips called chiplets, leading to higher scalability, easier integration, and better cost efficiency. This approach has recently become a research hot spot with several works investigating different aspects of chiplet design\cite{marvell-co-a,marvell-co-b}.
Besides the benefits of chiplets, however, there are also challenges in designing high-performance and low-cost chiplets. The interconnection network is one of them. Chiplets are usually connected through a shared network where different types of chiplets compete for limited network resources. Contention between the chiplets can lead to sub-optimal resource utilization and performance degradation if the network resource is not carefully managed.

In this work, we investigate the interconnection network design for systems consisting of CPU chiplets and GPU chiplets. These chiplets are connected through a shared network, as depicted in Figure~\ref{fig:cpu-gpu}. In this system, a tile can be a CPU-chiplet containing CPU cores or a GPU chiplet containing GPU cores. The interconnection network uses a mesh topology and is implemented in the interposer of the package. Each chiplet is connected to the network through a router that routes data packets generated by the chiplets.

In such a heterogeneous chiplet system containing both CPUs and GPUs, providing fair sharing of the network resources is challenging. Conventional CPU applications are latency-sensitive and have lower Thread Level Parallelism (TLP), while GPU applications are more sensitive to bandwidth because of their higher TLP\cite{Zhuren, amoeba-gpu}. Such diverse traffic patterns make appropriate sharing of resources a critical challenge in the network design \cite{lc-conflict-free,onur}. In this work, we develop a reconfigurable network architecture to improve resource utilization. Instead of employing a static method to resolve traffic contention between CPU and GPU packets, we leverage the Kalman Filter to predict each application's dynamic request for network resources. Then the network resource is dynamically allocated to each chiplets to meet their demand.

\section{Motivation}
\begin{figure}[t!]
    \centering
    \scalebox{1}{\includegraphics[width=0.9\linewidth]{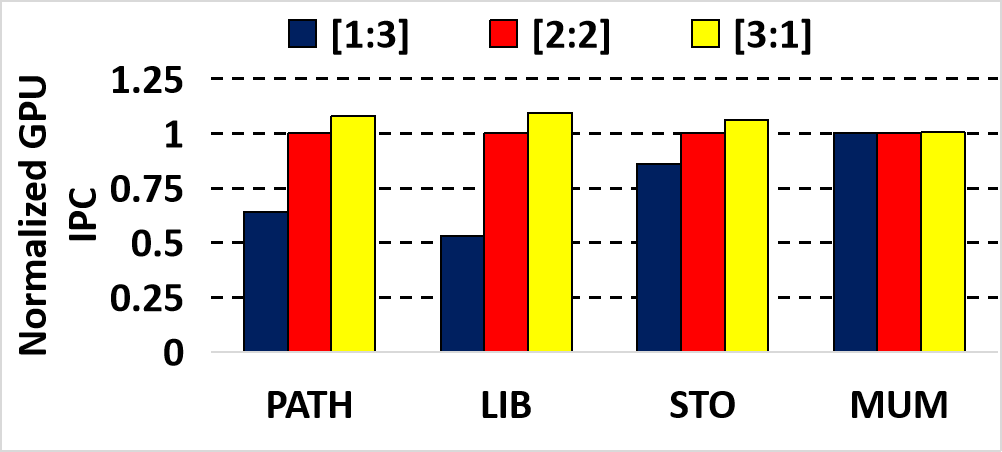}}
    \vspace{-0.2 cm}
 \caption{GPU performance with different VC allocation ratio between GPUs and CPUs [GPU VCs:CPU VCs].}
    \label{fig:sub1}
    \vspace{-0.2 cm}
\end{figure}

\begin{figure}[t!]
    \centering
    \scalebox{1}{\includegraphics[width=0.9\linewidth]{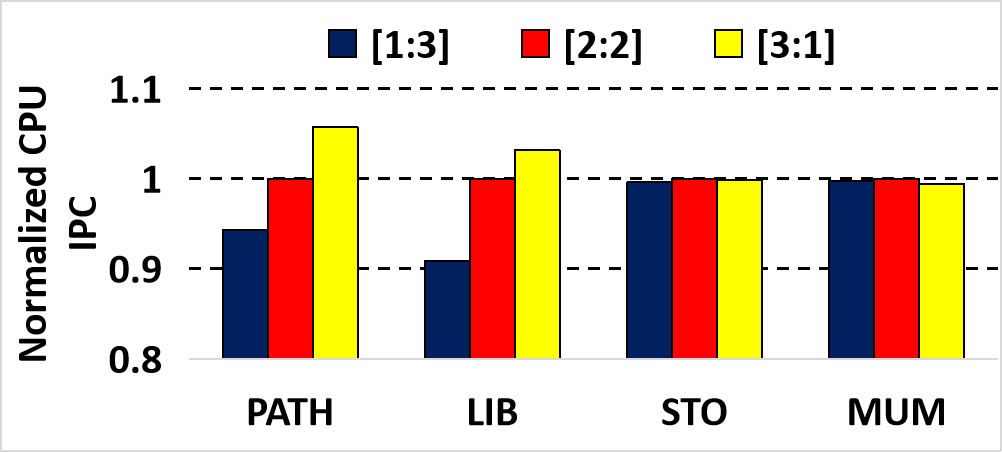}}
    \vspace{-0.2 cm}
 \caption{CPU performance with different VC allocation ratio between GPUs and CPUs [GPU VCs:CPU VCs].}
    \label{fig:sub2}
    \vspace{-0.5 cm}
\end{figure}

In heterogeneous computing systems, effective resource management not only needs to consider the performance optimization of computing units (such as GPUs and CPUs) but also needs to pay attention to the efficiency of the underlying communication network. Data transmission and processing are implemented through network routing, where the configuration of Virtual Channels (VCs) and Switch Arbitration (SA) plays a crucial role in reducing congestion, lowering latency, and enhancing data transfer efficiency. Static or simple dynamic allocation mechanisms are often used due to their low complexity. However, these mechanisms cannot fully adapt to rapid changes in computing requirements, resulting in poor performance in highly dynamic environments. Furthermore, because GPUs and CPUs have different computing characteristics, running these two types of applications together may lead to contention for shared network resources and result in suboptimal system performance. Therefore, it is vital to develop a resource management strategy that can accurately predict and dynamically adapt to these changes. 

\subsection{Virtual Channel Allocation}
We investigated the impact of VC allocation on the CPU and GPU performance by varying the number of VCs allocated to them. The network uses 4 VCs for each router input that are shared between CPU and GPU packets. For example, [1:3] configuration allocates 1 VC to GPU traffic, and 3 VCs to CPU traffic. In the heterogeneous chiplet system we investigated, all CPU chiplets run application \textit{omnetpp} while the GPU chiplets run one application from \textit{PATH, LIB, STO, and MUM} each time. Figure~\ref{fig:sub1} shows the GPU IPC with different partition schemes of VCs. As can be observed, GPU performance gets improved by allocating more VC resources. Figure~\ref{fig:sub2} shows the CPU IPC with different partitions of VCs. Allocating more resources to CPUs, on the other hand, does not result in significant performance improvement. In some cases, allocating more resources to CPUs can even result in performance degradation for both CPUs and GPUs. This is because a significant number of CPU packets pile up at memory controllers (MCs), which already have many GPU packets waiting to be served. Increased CPU packets exacerbate the interference between the CPU and GPU traffic, resulting in increased latency for both of them. As a result, both CPU and GPU chiplets suffer from performance degradation.

\subsection{Switch Utilization}
\begin{figure}[t!]
    \centering
    \scalebox{1.1}{\includegraphics[width=0.9\linewidth]{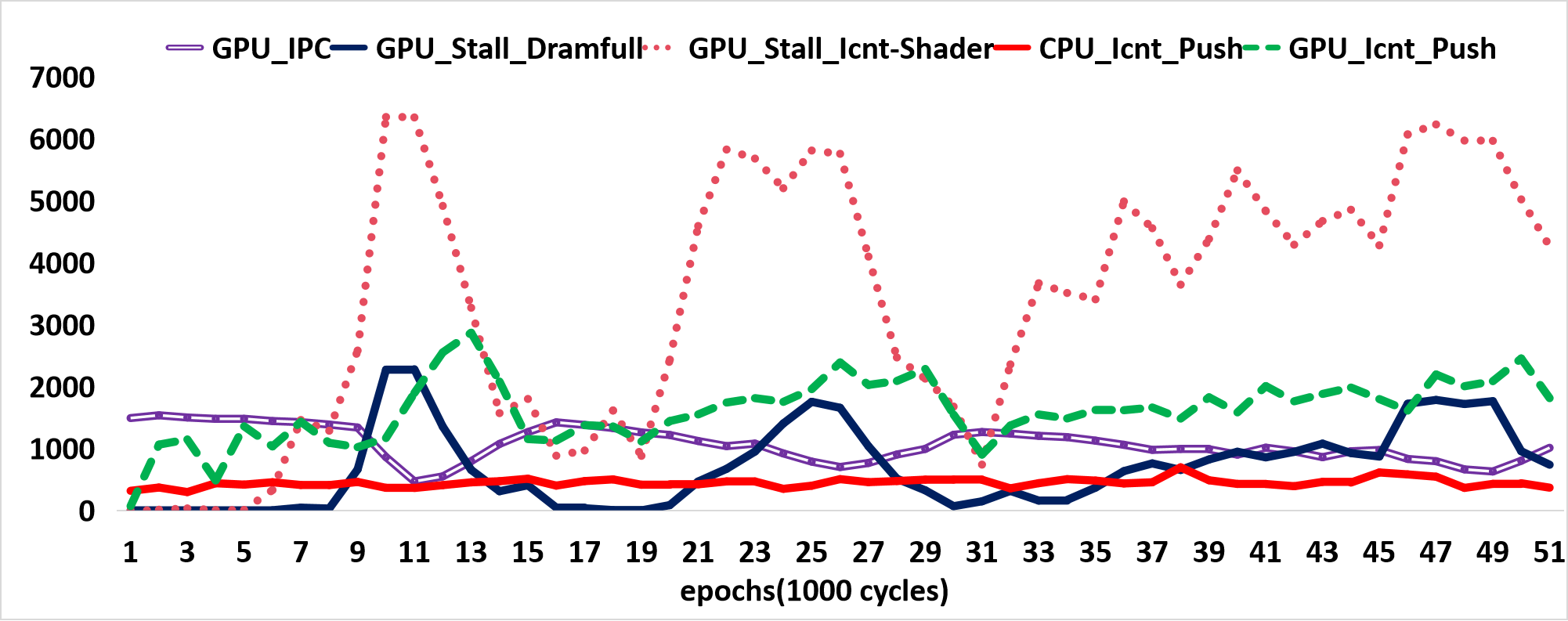}}
    \vspace{-0.4 cm}
 \caption{Dynamic Traffic Pattern of CPU and GPU chiplets.}
    \label{fig:MO3}
    \vspace{-.3in}
\end{figure}

In the microarchitecture of a NoC router, the switch allocation module is responsible for managing the switch traversal of data packets within the router to ensure that data packets can be forwarded efficiently from their input ports to their output ports. SW arbitration is performed among multiple packets competing for the same output port. The winner of this arbitration can move to the next hop while other packets have to wait for the next cycle's arbitration.  
From Figure~\ref{fig:MO3}, we can observe that GPU traffic injection rates change significantly in different time periods, while CPU traffic injection is relatively stable. Considering the variation of GPU traffic with time, dynamically adjusting the switch arbitration strategy to allocate more switch utilization to GPUs can allow their packets to move faster and thus alleviate network congestion.
\nopagebreak[3]
\section{Reconfigurable Network Design Using Kalman Filter}

\begin{figure}[t]
    \centering
    \scalebox{1}{\includegraphics[width=0.9\linewidth]{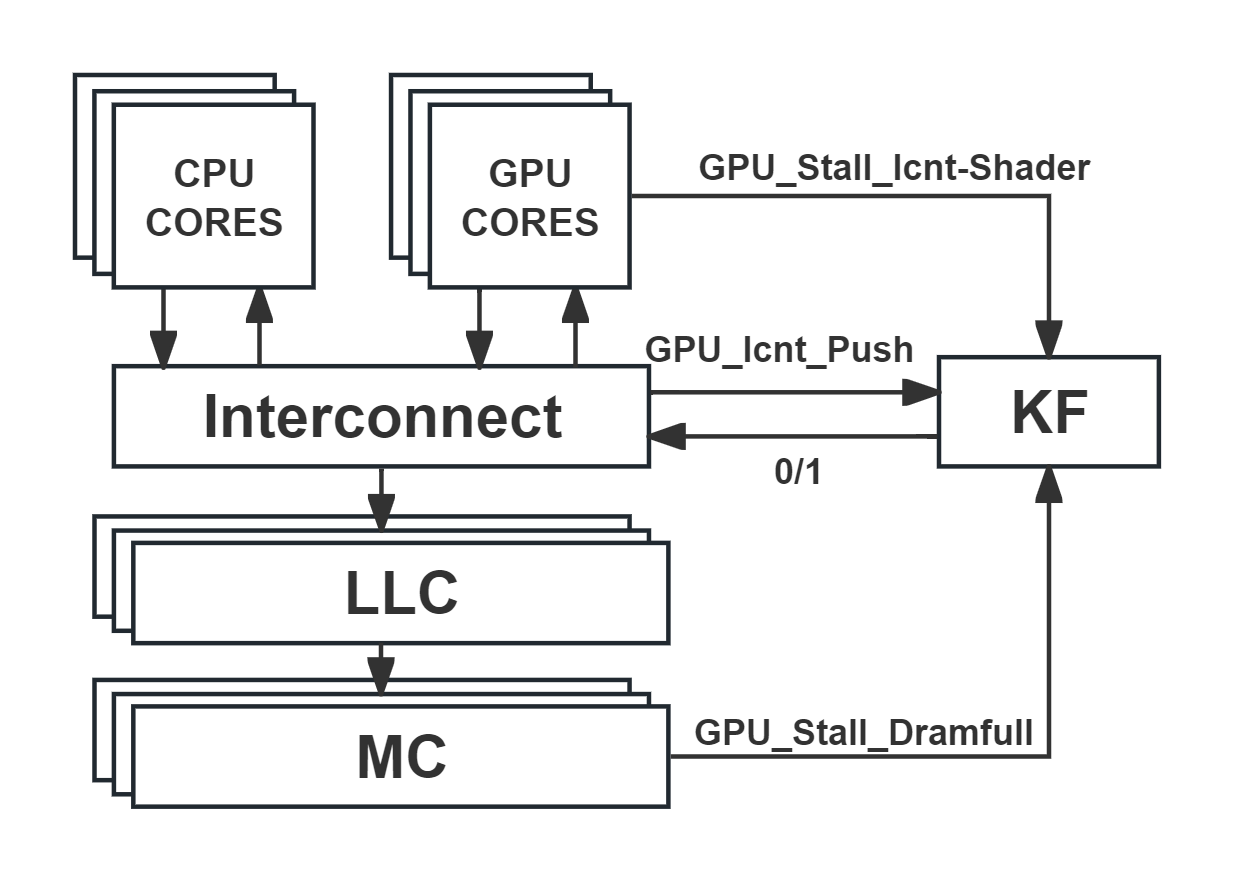}}
    \vspace{-0.6 cm}
 \caption{System Overview.}
    \label{fig:arch1}
    \vspace{-0.6 cm}
\end{figure}

As an effective and low-cost predicting technique, Kalman Filter (KF) offers a viable solution for embedded systems \cite{kalman}. By leveraging historical data and real-time feedback, KF can accurately predict future changes in system states, providing support for better resource allocation decisions. This motivated us to develop a KF-based scheme to dynamically allocate network resources in heterogeneous chiplets, aiming for better performance at lower cost. We investigated shared network resources and found two critical resources the applications competing for which majorly determine the performance: virtual channel (VC) and switch utilization. 
\subsection{Kalman Filter Algorithm}

Kalman Filter is an efficient recursive filter, which can estimate the system state and its error covariance even if the transition and observation models are uncertain.  It utilizes two sets of equations: time update equations and measurement update equations. Initially, KF employs time updates to predict the current state estimation ahead of time. Afterward, it updates these estimations through the measurement equations, incorporating real-time measurements to correct the forecasts. The time update equation works as a predictor, predicting the system's future state, whereas the measurement update equation acts as an adjuster, refining these predictions with observed data.

The KF implements a feedback control mechanism for state estimation by utilizing the measurements as feedback. It advances the current state and error covariance through the time update equation to predict a priori estimates of the next state. By integrating new measurements with past estimates via the measurement equations, it refines these estimates and thus improves posterior accuracy. This process balances prediction and correction to continuously update its understanding of the system state.

\(X_{k-1}\) is the prior state estimation, \(\hat{X}_{k}\) and \(X_{k}\) are predictive and posteriori state estimation. \(A\), \(B\) and \(H\) are state transition, control-input, observation model, \(Q\) and \(R\) are covariance of process and observation noise. \(P_{k-1}\) is the prior estimation error covariance, \(\hat{P}\) and \(P_{k}\) are predictive and posteriori estimation error covariance. \(U\) and \(I\) are control and unit matrix. The formula for implementing a Kalman Filter is described below: 

\textbf{Kalman Filter Time Update Equations:}
\begin{align*} &\hat{X}_{k}= AX_{k-1}+ BU_{k-1} \tag{1}\\ &\ \ \hat{P}_{k}= AP_{k-1}A^{T}+Q \tag{2} \end{align*}
The state and covariance estimates from time step k-1 to step k can be obtained by Eqs.(1) and Eqs.(2).

\textbf{Kalman Filter Measurement Update Equations:}
\begin{align*} &K_{k}=\hat{P}_{k}H^{T}(H\hat{P}_{k}H^{T}+R)^{-1} \tag{3}\\ &\ \ X_{k}=\hat{X}_{k}+ K_{k}(Z_{k}-H\hat{X}_{k}) \tag{4}\\ &\qquad\ P_{k}=(I- K_{k})\hat{P}_{k} \tag{5} \end{align*}
In the measurement update phase, the process begins by calculating the Kalman gain \(K_{k}\), as shown in Eqs.(3). The posterior state estimate is then obtained by combining the measurements using Eqs.(4), an iterative formula that converges to the final estimate after several iterations. Finally, the update phase recalculates the error covariance for each iteration, as specified in Eqs.(5), ensuring the estimates are adjusted for the actual measurements received.

\begin{figure}[t]
    \centering
    \scalebox{1}{\includegraphics[width=0.9\linewidth]{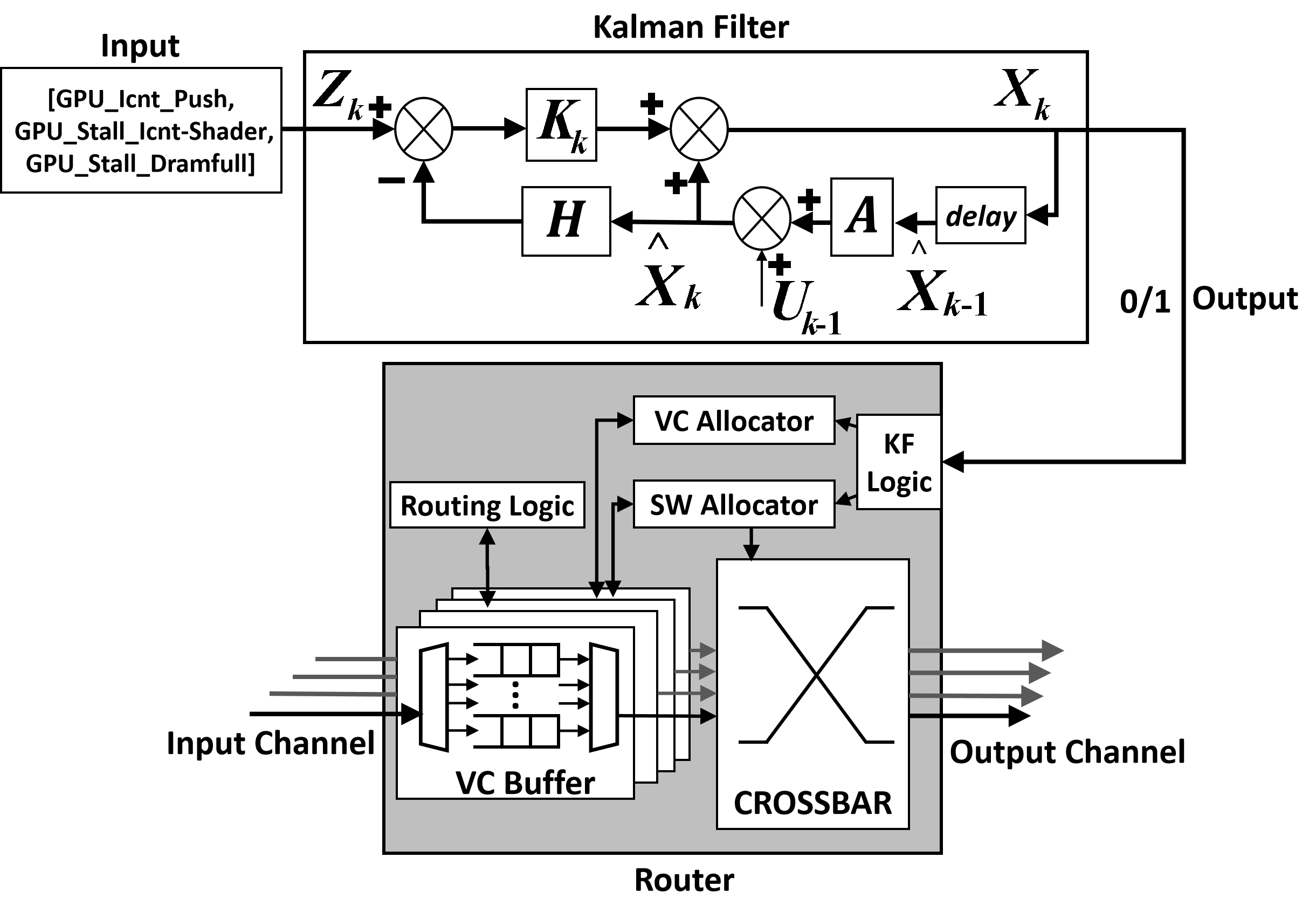}}
    \vspace{-0.2 cm}
 \caption{Kalman Filter and Router MicroArchitecture Design}
    \label{fig:arch2}
    \vspace{-0.6 cm}
\end{figure}

\subsection{Predicting NoC Traffic through Kalman Filter}
As can be observed from Figure~\ref{fig:MO3}, there exists a notable correlation between GPU IPC and these three metrics::
\begin{itemize}
    \item \texttt{GPU\_Icnt\_Push}: The number of injections from the GPU core to the Interconnection Network (ICNT).
    \item \texttt{GPU\_Stall\_Icnt-Shader}: The number of GPU stalls caused by delays from the ICNT to shader cores.
    \item \texttt{GPU\_Stall\_Dramfull}: The number of GPU stalls due to DRAM memory being full.
\end{itemize}
When GPU\_Icnt\_Push increases, which leads to more network congestion, GPU\_Stall\_Icnt-Shader and GPU\_Stall\_Dramfull will increase, leading to a drop in IPC. This observation enables our construction of a Kalman Filter model that focuses on GPU-related data because the CPU-related metrics barely change. To handle the significant variance among these metrics, we first preprocess the data and normalize them, so that they fall into the specific interval of [-1, 1]. The purpose of this standardization of data is to scale the data to a reasonable range, which can help improve the efficiency and accuracy of the model in processing the data.

In our KF design, the state variable, denoted as \(X_{k}\), represents the GPU's IPC, which we want to predict. Our dynamic network parameters include: GPU\_Stall\_Dramfull, GPU\_Icnt\_Push, GPU\_Stall\_Icnt-Shader, the observations are denoted as vectors \(Z_{k}=[Z_{k1},Z_{k2},Z_{k3}]^{T}\). Through these parameter settings, the KF can predict the GPU’s IPC in the next epoch. When the KF output is negative, it means that the GPU's IPC is maintained at a high level. When it becomes positive, it means that the GPU's IPC will decline. We can dynamically adjust VC allocation and Switch Arbitration to make the GPU get more resources. Figure~\ref{fig:arch1} shows the architectural integration of a Kalman Filter within heterogeneous chiplet systems. GPU and CPU cores are linked to a network. GPU\_Stall\_Icnt-Shader is sourced from data flow management between GPU cores and the interconnection network. GPU\_Icnt\_Push comes from the interconnection network, which manages data transfer between the GPU cores and other components. GPU\_Stall\_Dramfull is retrieved from the memory controllers (MCs) that manage the DRAM. This interface acts as a gateway for feeding each epoch’s data into the KF module. Figure~\ref{fig:arch2} shows more details of a router's microarchitecture design. The Kalman Filter's output directs VC and SW Allocator adjustments via the KF logic, optimizing network resource allocation and system performance by dynamically reacting to congestion and traffic conditions.

\begin{figure}
    \centering
    \scalebox{0.9}{\includegraphics[width=0.9\linewidth]{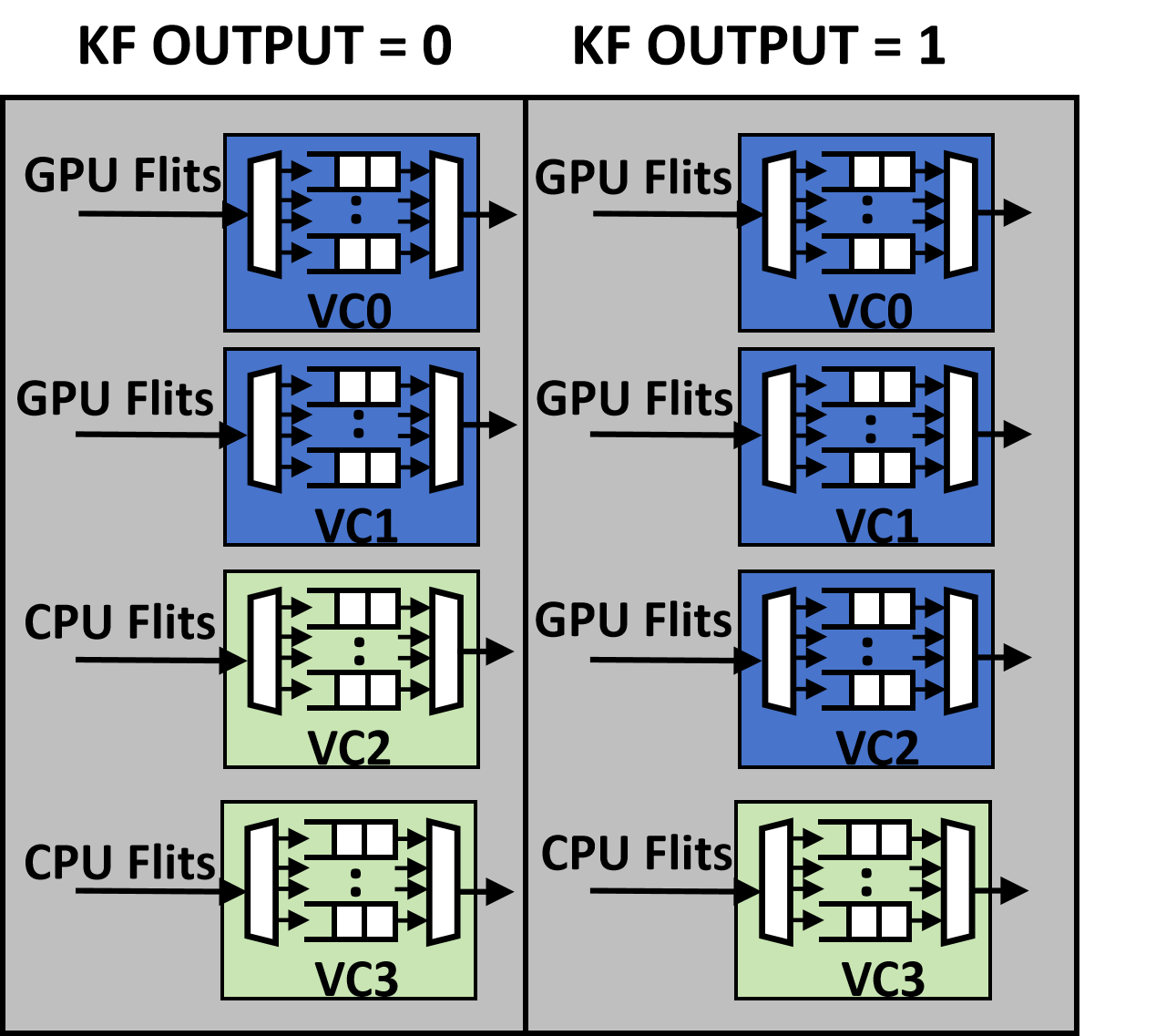}}
    \vspace{-0.2 cm}
 \caption{Reconfigurable VC Allocation.}
    \label{fig:vcallocate}
    \vspace{-0.2 in}
\end{figure}

To further optimize our approach, we also formulated the rules for deploying the Kalman Filter within our system. Initially, network resources are distributed equally between the GPUs and CPUs. To maintain the system stability and avoid deadlocks, the KF is not activated until 10,000 cycles after the GPU applications start running. After any network resource reallocation, the new configuration is maintained for a minimum of 5,000 cycles. Therefore, even with a change in the KF's predictive output, adjustments to resource allocation are deferred until after the mandated period has elapsed. In situations where the duration in the state (KF output=1) surpasses 10,000 cycles, a reduction in GPU resource allocation might be advisable, aiming for a return to an equitable sharing of network resources between the CPUs and GPUs.

\subsection{Dynamic VC and Switch Allocation}
Our baseline method of VC allocation uses iSLIP\cite{769767} which allows for efficient, round-robin scheduling. Our approach is a reconfigurable network design that segregates flits into GPU and CPU partitions, allocating private VCs to each specific node that is critical to synchronized CPU and GPU operations. This strategy assigns a first set of VCs to CPU traffic and the remaining set of VCs to GPU traffic, ensuring dedicated resources for each type of traffic flow. This can reduce congestion and therefore improve system performance. Allocations are dynamically adjusted based on the VC index associated with each flit type. As illustrated in Figure~\ref{fig:vcallocate}, when the KF output is 0, GPU traffic uses VC 0 and 1, and CPU traffic uses VC 2 and 3. When the KF output shifts to 1, GPU traffic can use VC 0 through 2, while CPU traffic can only go through VC 3.

\begin{figure}
    \centering
    \scalebox{0.9}{\includegraphics[width=0.9\linewidth]{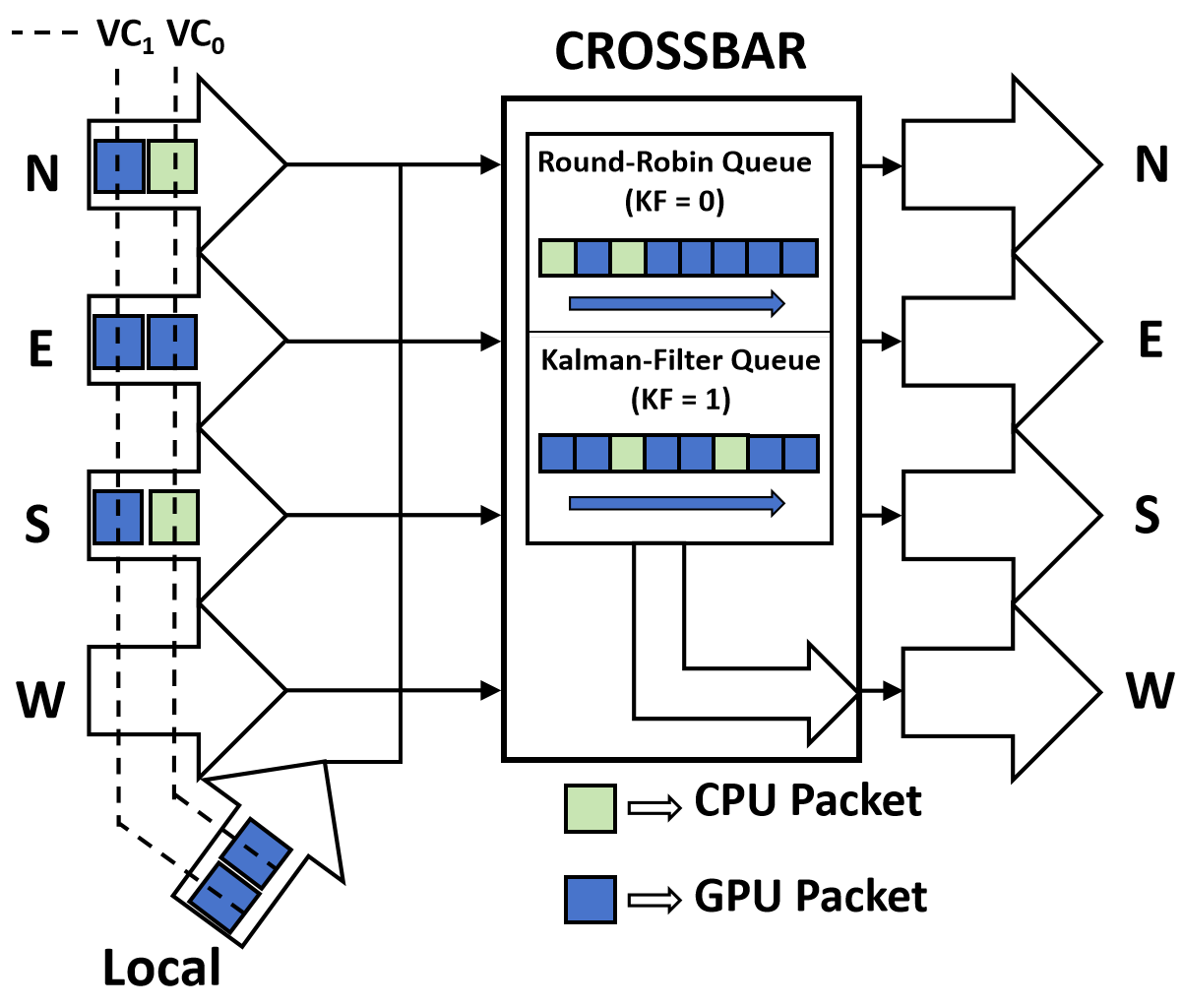}}
    \vspace{-0.2 cm}
 \caption{Reconfigurable Switch Allocation.}
    \label{fig:swallocate}
    \vspace{-0.2 in}
\end{figure}

Figure~\ref{fig:swallocate} illustrates an example of the switch output request queue in our routers. We use a single output request queue to explain our switch allocation policy triggered by Kalman Filter (KF) prediction. When a packet arrives at any input port of a router, it requests access to an output port to proceed to its destination. In cases where multiple input ports request the same output port, the switch allocator must decide which input port to grant access to in every cycle. The baseline switch allocation utilizes a round-robin policy, as depicted in Figure~\ref{fig:swallocate}, where blue blocks represent GPU packets and green ones represent CPU packets. The baseline routers employ a round-robin policy, each packet is sent to the output port in a round-robin manner, resulting in the first queue shown in the figure. In our design, if the KF output is "0", the round-robin policy will take effect. In this example, CPU packets will be granted access to the output port faster than GPU packets due to a round-robin policy. When KF predicts the need to allocate more resources to GPU packets (KF output = 1), we require a different starvation-free policy to prioritize GPU packets over CPU packets. Our new policy prioritizes two GPU packets followed by one CPU packet. In this way, more switch bandwidth resource is allocated to GPUs, trying to meet their increased demand for network resources.

\section{Evaluation}
\subsection{Experiment Setup}
We use the GPGPU-sim simulator\cite{gpgpusim} and integrate it with a cycle-level x86 CMP simulator to evaluate our heterogeneous chiplet system. Our baseline configuration is shown in Table~\ref{tab:system_configurations}. We experimented with GPU benchmarks from the ISPASS2009\cite{gpgpusim} and Rodinia \cite{rodinia}. The CPU benchmarks are selected from the SPEC CPU 2006 INT and FP suites and commercial server workloads for CPU. In our baseline, we employed two subnets: one for CPU/GPU requests and one for CPU/GPU reply packets to avoid protocol deadlocks. The baseline allows CPU and GPU chiplets to share network and router resources in a round-robin fashion. Additionally, for comparison purposes, we implemented a fair partition of network resources, equally allocating the resources between CPU and GPU packets. For more comparison, we have also implemented four subnets, physically segregating CPU and GPU packets, allocating a dedicated subnet for CPU requests and replies, as well as GPU requests and replies.

\begin{table}
    \centering
    \caption{System Configurations}
    \label{tab:system_configurations}
    \small
    \begin{tabular}{|p{1.5cm}|p{6.5cm}|}
        \hline
        \textbf{GPU} & 14 Chips = 28 SM, 700MHz  \\
        \hline
        \textbf{GPU Spec} & Max 1536 Threads (48 warps, 32          threads/warp), 64KB Shared Memory, 64KB Register \\
        \hline
        \textbf{GPU Caches} & 16KB 4-way (1 Data Cache), 12KB 24-way     texture, 8KB 2-way Constant Cache, 2KB 4-way I-Cache, 128B Line     Size \\
        \hline
        \textbf{CPU} & 14 x86 Cores, 2000MHz, 128-entry instruction window,     OoO Fetch \& Execute, 3 instruction/cycles,max, 1 memory            instructions/cycle \\
        \hline
        \textbf{CPU L1 Cache} & 32KB 4-way, 2-cycle lookup, 128B Line Size \\
        \hline
        \textbf{CPU L2 Cache} & 256KB 8-way, 8-cycle lookup, 128B        Line Size \\
        \hline
        \textbf{Shared LLC} & 1 MB/Memory Partition, 128B Line, 16-       way, 700MHz \\
        \hline
        \textbf{Warp Scheduler} & Greedy-then-oldest\\
        \hline
        \textbf{Features} & Memory Coalescing, Inter-warp Merging,              Interconnect \\
        \hline
        \textbf{Interconnect} & 6 × 6 Shared 2D Mesh, 1400MHz, XY Routing,      2 GPU cores per node, 1 CPU core per node, 32B Channel Width, 16VCs, Buffers/VC = 4\\
        \hline
        \textbf{Memory Controller} & 8 Shared GDDR5 MCs, 800 MHz, FR-FCFS,      8 DRAM-banks/MC\\
        \hline
    \end{tabular}
\end{table}

\begin{figure}[t!]
    \centering
    \scalebox{1.1}{\includegraphics[width=0.9\linewidth]{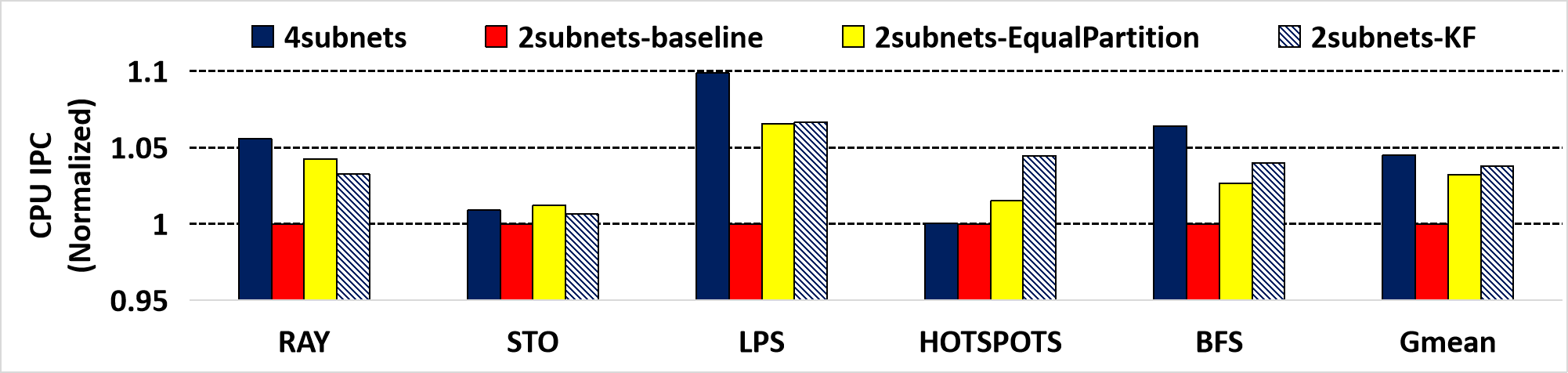}}
    \vspace{-0.2 cm}
 \caption{CPU IPC.}
    \label{fig:cpu-ipc}
    \vspace{-0.2 cm}
\end{figure}

\begin{figure}
    \centering
    \scalebox{1.1}{\includegraphics[width=0.9\linewidth]{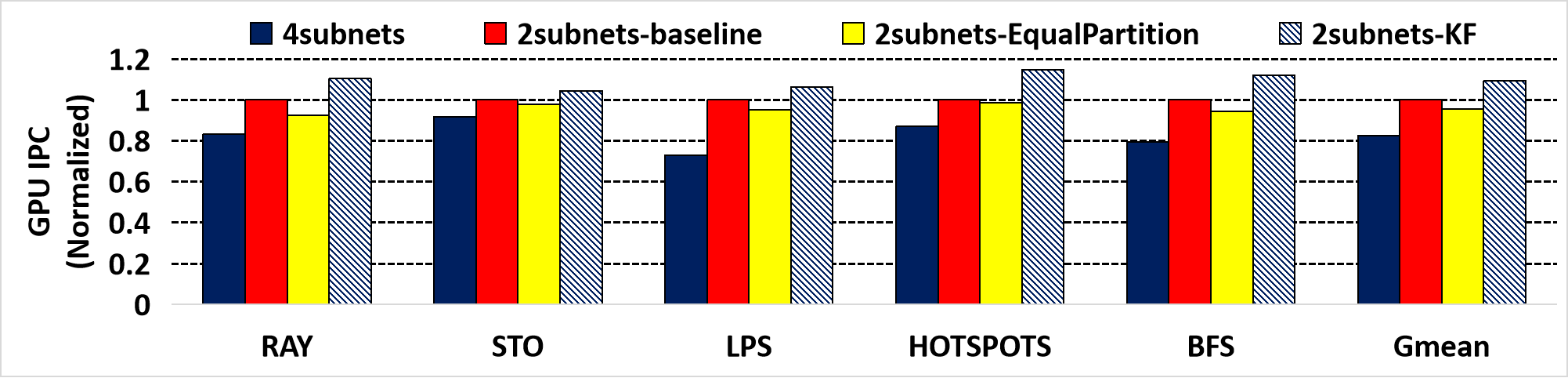}}
    \vspace{-0.2 cm}
 \caption{GPU IPC.}
    \label{fig:gpu-ipc}
    \vspace{-0.2 cm}
\end{figure}

\begin{figure}
    \centering
    \scalebox{1.1}{\includegraphics[width=0.9\linewidth]{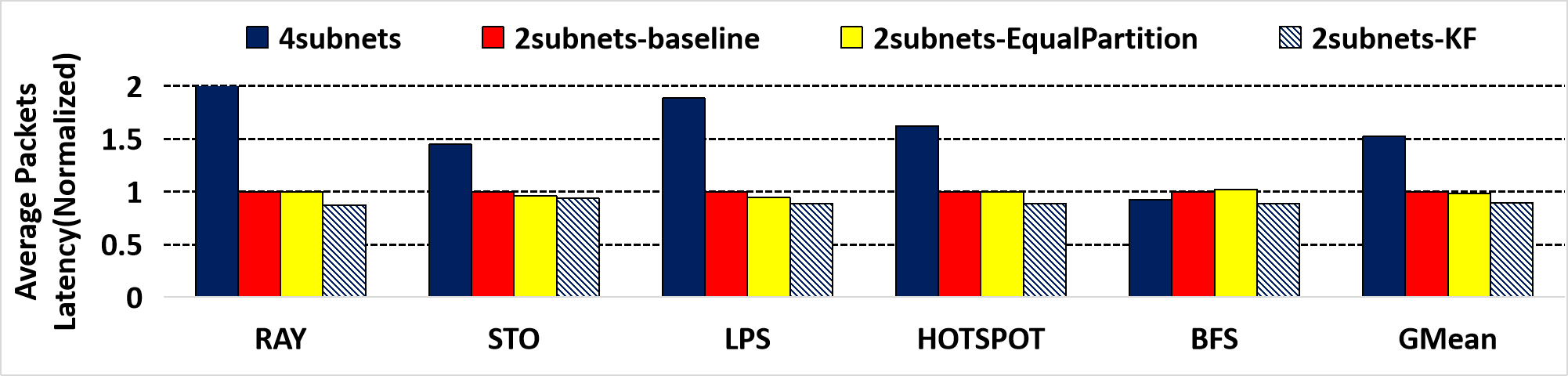}}
    \vspace{-0.2 cm}
 \caption{Average Packet Latency.}
    \label{fig:latency}
    \vspace{-0.2 cm}
\end{figure}

\begin{figure}
    \centering
    \scalebox{1}{\includegraphics[width=0.9\linewidth]{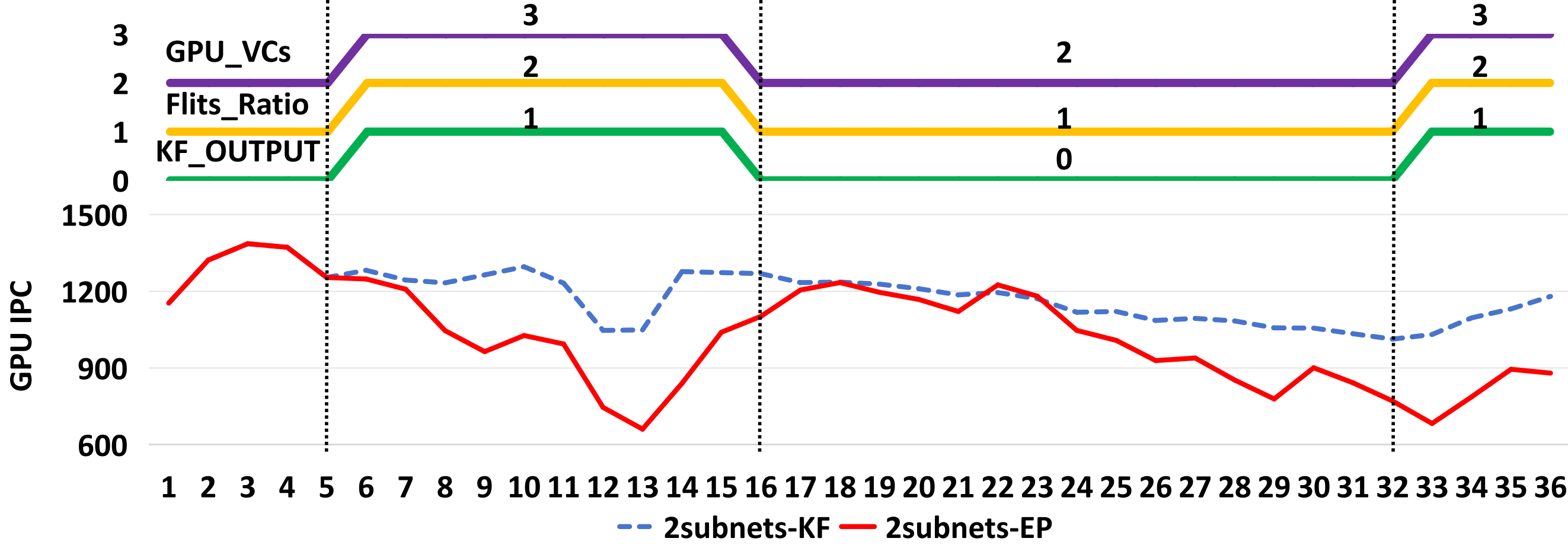}}
    \vspace{-0.2 cm}
 \caption{Dynamic GPU Performance with and without Kalman Filter-assisted Resource Allocation.}
    \label{fig:gpu-epoch}
    \vspace{-0.2 cm}
\end{figure}
\subsection{Result Analysis}
We evaluated four different configurations, as depicted in Figures~\ref{fig:cpu-ipc} and~\ref{fig:gpu-ipc}. The first configuration uses four subnets, with two separate subnets allocated for CPU and GPU traffic (one for requests and one for reply packets). The second configuration, serving as the baseline, employs two subnets and utilizes round-robin scheduling to share network resources, which has the advantage of reduced cost and employs a fair scheduling policy for on-chip routers. The third configuration equally distributes network resources, with each GPU and CPU flow having an equal number of VCs allocated. 
The last configuration is for our reconfigurable network design, where we utilized Kalman Filtering to predict the next epoch traffic pattern and adjust the resource allocation accordingly.

We can observe from Figure~\ref{fig:latency} that the packet latency is higher when using four subnets compared to other designs with two subnets. This is because the four exclusive subnets do not allow sharing between different traffic flows and therefore cannot fully utilize the network bandwidth. In the case of using only two subnets, we allocated half of the VCs for the reply subnet, allowing GPUs to take over bandwidth not used by the CPUs. As shown in Figure~\ref{fig:latency}, the performance of fair partitioning is similar to the baseline with a small reduction in STO and LPS. In comparison, our proposed Kalman Filter-based approach reduces the packet latency across all of our workloads.

The main reason for this reduction in packet latency is better resource management. Figure~\ref{fig:MO3} illustrates that in certain epochs when the GPU chiplets inject more packets, the number of GPU stalls increases. This is primarily due to the imbalanced resource allocation when the GPU needs more resources but cannot get them from the CPU chiplets. This imbalance results in performance degradation of GPUs. Figure~\ref{fig:gpu-epoch} shows that in some epochs, the two subnets with equally distributed resources (2-subnet-fair) experience a drop in IPC which means this resource allocation method cannot adjust to dynamic traffic changes. Using our proposed technique, by leveraging the Kalman Filter, we can predict these dynamic changes accurately. Then our resource allocator can adjust the allocation policy and allocate more resources to GPUs to avoid their performance degradation. 

Figure~\ref{fig:gpu-ipc} shows the GPU IPC. The 4-subnet configuration experienced a decrease in performance across all five workloads, with nearly a 20\% reduction for BFS. In the 2-subnet fair allocation scheme, we observed a slight decrease in IPC compared to the baseline. This is due to the higher bandwidth requirement by the GPU in the network, as GPU applications are typically more memory-intensive than CPU applications and require more network resources to meet their demands. On the other hand, our KF-based dynamic allocation predicted correctly the epochs when GPU cores need more resources. By detecting an increase in packet injection and longer GPU stalls due to memory accesses, as well as GPU DRAM-full stalls, as depicted in Figure~\ref{fig:MO3}, KF can predict increased resource demand from GPUs. Accordingly, we can adjust the resource allocation policy and reduce the GPU stall time. As a result, GPU IPC increased by up to 19\%. This shows our scheme can lead to more efficient resource sharing. As can be observed in Figure~\ref{fig:cpu-ipc}, our design can keep CPU performance unaffected and improve GPU performance by 7\%.
In Figure~\ref{fig:gpu-epoch}, the green line at the top represents the output signal generated by our KF predictor. A signal value of "0" indicates an equal distribution of resources is good enough, while a value of "1" indicates a change will be needed in the next epoch. For example, we can move from the equal partition of VCs between the CPUs and GPUs to giving more resources to GPUs, by allocating 75\% of resources to the GPU and 25\% to the CPU. This adjustment is implemented in both VC allocation and switch arbitration. 

\section{Related Work}

Prior works have investigated different aspects of accelerator noc desis. Wang et al. \cite{applicatio-defined}, studied chiplet-based SoC design, utilizing advanced packaging to integrate multiple chips.
Their technique can improve network bandwidth and reduce latency for AI applications. Cheng et al. proposed a novel noc architecture to remove performance bottleneck \cite{packetpump}. An optimized tree topology was proposed for GPU applications \cite{ancs}.  Vivet et al. \cite{intAct} tackled the challenges faced in high-performance computing due to the increasing complexity and costs associated with integrating various computing capabilities, including generic cores and AI accelerators. They highlighted the shift towards alternative architecture solutions such as chiplet-based systems using 3D technologies to achieve modular and scalable designs. Feng et al. \cite{scalable-methodology} proposed an interconnection method for chiplet-based systems to establish high-radix interconnection networks, overcoming the limitations of flat topologies like 2D-mesh and ensuring deadlock-free routing. 
\section{Conclusion}

The increasing cost and design complexity of large System-on-Chips (SoCs) have motivated the design of chiplets. The increasing demand for heterogeneous hardware architectures leads to the development of heterogeneous chiplets. However, chiplet design faces several challenges and the interconnection network is a prominent one. 
In this work, we propose a reconfigurable chiplet interconnection design to improve resource utilization by leveraging Kalman Filter. Our technique can dynamically adjust the network resources allocated to CPU and GPU chiplets based on their needs predicted by Kalman Filter. Our technique can improve both CPU and GPU performance through efficient resource allocation.

\begin{acks}
This work was supported in part by NSF grants 2008911, 2046186, and 2051062.
\end{acks}


\bibliographystyle{ACM-Reference-Format}
\bibliography{refs}

\end{document}